\documentclass[%
 reprint,
nofootinbib,
 amsmath,amssymb,
 aps,
]{revtex4-2}

\usepackage[usenames]{color} 
\usepackage{amssymb} 
\usepackage{amsmath} 
\usepackage{float}
\usepackage[pdftex]{graphicx}
\usepackage[utf8]{inputenc} 
\usepackage[colorlinks=true,urlcolor=blue,linkcolor=blue,citecolor=blue]{hyperref}       
\usepackage[bottom]{footmisc}
\usepackage{multirow}

\usepackage{dcolumn}
\usepackage{slashed}
\usepackage{physics}
\usepackage{wrapfig}
\usepackage{lineno}
\usepackage{multirow}
\usepackage{xspace}
\usepackage{booktabs}
\usepackage{tipa}
\usepackage{bm}
\usepackage{tabularx} 
\usepackage{adjustbox}

\newlength\cmsTabSkip
\newcommand{\ptvecmiss}{\ensuremath{{\vec p}_{\mathrm{T}}^{\kern1pt\text{miss}}}\xspace}

\begin{document}




\title{
Re-Simulation-based Self-Supervised Learning for Pre-Training Physics Foundation Models}

\author{P. Harris}
\affiliation{Massachusetts Institute of Technology, Cambridge, United States}
\affiliation{Institute for Artificial Intelligence and Fundamental Interactions, Cambridge, United States}
\author{M. Kagan}
\email{makagan@slac.stanford.edu}
\affiliation{SLAC National Accelerator Laboratory, Stanford, United States}
\author{J. Krupa}
\email{jkrupa@mit.edu}
\affiliation{Massachusetts Institute of Technology, Cambridge, United States}
\affiliation{Institute for Artificial Intelligence and Fundamental Interactions, Cambridge, United States}
\affiliation{SLAC National Accelerator Laboratory, Stanford, United States}
\author{B. Maier}
\email{b.maier@imperial.ac.uk}
\affiliation{I-X, Imperial College, London, United Kingdom}
\author{N. Woodward}
\affiliation{Massachusetts Institute of Technology, Cambridge, United States}

\begin{abstract}
Self-Supervised Learning (SSL) is at the core of training modern large machine learning models, providing a scheme for learning powerful representations that can be used in a variety of downstream tasks. However, SSL strategies must be adapted to the type of training data and downstream tasks required. We propose  RS3L (``Re-simulation-based self-supervised representation learning''), a novel simulation-based SSL strategy that employs a method of \emph{re-simulation} to drive data augmentation for contrastive learning in the physical sciences, particularly, in fields that rely on stochastic simulators. By intervening in the middle of the simulation process and re-running simulation components downstream of the intervention, we generate multiple realizations of an event, thus producing a set of augmentations covering all physics-driven variations available in the simulator. Using experiments from high-energy physics, we explore how this strategy may enable the development of a foundation model; we show how RS3L pre-training enables powerful performance in downstream tasks such as discrimination of a variety of objects and uncertainty mitigation. In addition to our results, we make the RS3L dataset publicly available for further studies on how to improve SSL strategies.
\end{abstract}

\maketitle

\newpage

\section{Introduction}
\label{sec:intro}

Self-Supervised Learning (SSL) has been a key driver in recent machine learning (ML) advancements;
through the use of data augmentation, SSL algorithms rely on pseudo labels to create representations of the data that can be highly useful and efficiently adapted for downstream tasks. Accordingly, the resulting SSL trained model can act as a powerful \emph{foundation model}~\cite{bommasani2022opportunities}, defined here as a model that is ``pre-trained'' on a generic task, in this case one defined with SSL, and capable of being fine-tuned for a variety of purposes. Foundation models have quickly arisen as a strategy for describing complex tasks, especially in computer vision (e.g.,~\cite{tong2022videomae}), natural language processing (e.g.,~\cite{brown2020language}), and speech recognition (e.g.,~\cite{yang2021superb}).  More recently, foundation models for machine learning in science have become an important and active area of research, for instance in biology~\cite{ESM}, chemistry~\cite{LLM_chem1,LLM_chem2,Irwin_2022,ahmad2022chemberta2}, differential equation solving~\cite{subramanian2023foundation,mccabe2023multiple}, cosmology~\cite{lanusse2023astroclip,walmsley2022galaxy}, and, finally, high-energy physics~\cite{Birk:2024knn}, not all of which, however are necessarily SSL-based.

SSL benefits from unlabeled training and from the creation of a relation between data augmentations. Leveraging vast unlabeled datasets can enable SSL to outperform supervised learning.
In this work, we focus on contrastive learning, a variant of SSL. We propose a new strategy for data augmentation and study how it performs within the SimCLR framework, which was first introduced in~\cite{DBLP:journals/corr/abs-2002-05709}. In SimCLR, the learning objective is to map a data point and its augmentation(s) to similar representations, while pushing different data points toward differing representations. In contrastive learning, the quality of the learning task is highly dependent on the set of augmentations. Domain completeness, i.e., when the augmentation set covers all possible variations in the data, can lead to better representation learning because the model can gain a full view of plausible data during learning.

To that end, we propose a simulation-based strategy to obtain an augmented data set that is as domain complete as possible and only limited by the knowledge encapsulated in a simulator. More specifically, in settings with stochastic simulators, we propose a strategy whereby we \emph{intervene} in the middle of a simulation process, fix the upstream generated latent state of an event, and \emph{re-simulate} downstream components to sample augmentations from the set of all plausible observations for a given latent state. From a programmatic standpoint, the first step fixes all initial conditions and stochastic latent variables in the simulation up to the intervention point, while the second step re-samples the stochastic latent variables in the simulation after the intervention point and thus generates a new stochastic output of the simulation conditioned on the fixed latent state at the intervention point. Hence, this allows us to interpret augmentations in self-supervised learning in a novel way: the point at which we intervene in the simulation chain is the point at which further information downstream is considered an augmentation, above is information. The goal is then to learn a robust representation that contains as much information as possible while integrating out the variability due to the augmentations. We dub our new strategy RS3L (\textipa{["rIz\super{9}l]}, ``Re-simulation-based self-supervised representation learning''). We will show that this method allows for powerful contrastive pre-training that, through altering the simulator settings for re-simulation, also aids in the mitigation of uncertainties arising from domain shift between potentially imperfect simulations and real data.

To test our method, we will focus our experiments on High Energy Physics (HEP), where high-fidelity simulators are readily available (see~\cite{Campbell:2022qmc} for an overview). In our case, the fixed latent state is represented by the elementary particle generated in a hard-scattering process, which can be described with perturbative quantum field theory. We then re-simulate the steps in which secondary particles are created through radiation off of the particles produced in the hard scattering, the subsequent hadronization of secondary particles into stable, bound particle states, and then the stable particle interaction with the detector material. In the following, this step is referred to as \emph{parton showering}. 

In particular, we will focus on learning representations of jets, which are highly energetic, collimated streams of secondary particles. 
Jets are core objects in HEP; they are produced from the showering and hadronization of high-energy quarks and gluons or through the decay of heavy elementary particles like the Higgs boson, the top quark, or heavy weak gauge bosons. Jets are featured in many theories for new physics beyond the Standard Model. Hence, the ability to ``tag'' a jet, i.e., to correctly and robustly classify the particle which initiated the jet, is crucial for the success of the physics program of experiments like ATLAS~\cite{ATLAS:2008xda} and CMS~\cite{CMS:2008xjf} at the Large Hadron Collider (LHC)~\cite{Lyndon_Evans_2008} and future high energy collider experiments. For instance, one of the canonical classification tasks in HEP and a prerequisite for the precise understanding of Higgs boson couplings to bottom quarks~\cite{ATLAS:2018kot,CMS:2018nsn,ATLAS:2020jwz,ATLAS:2020fcp,CMS:2023vzh} is the discrimination between jets originating from Higgs bosons and jets originating from QCD. Our choice of re-simulating the parton shower is motivated by a desire to characterize the progenitor of a jet, while capturing the large variability that results for each parton shower. That is, the re-simulated shower defines an augmentation to the existing shower that we claim captures all the physics that we desire downstream.

By sampling from our high-fidelity simulator to obtain augmented versions of jets, our work presents an evolution of the methods developed in JetCLR~\cite{10.21468/SciPostPhys.12.6.188} and~\cite{Witkowski:2023xmx}, which focused on empirical methods of augmentation generation (such as rotations), whereas our work develops a domain-complete augmentation set based concretely on the physics in the simulation. As a result, we outline a strategy through augmentations within SSL to pre-train a latent space that captures the most salient, physically motivated, features of a simulation. Within high energy physics, other SSL methods have also been explored for foundation model pre-training. For instance, the masked modeling strategies developed in BERT~\cite{devlin2019bert} and BEiT~\cite{bao2022beit} have been adapted for set-type data in order to mask and predict particles in a jet~\cite{heinrich2024masked}, and to mask and predict particle-type information~\cite{kishimoto2023pretraining}. Supervised pre-training and fine-tuning strategies have also been explored for high energy physics contexts~\cite{supervisedfinetuning}. A similar strategy has been also pursued within astrophysics~\cite{Akhmetzhanova:2023hiy}.

Accordingly, the contributions of this paper are as follows:

\begin{itemize}
    \item The development of a methodology for re-simulation through intervening in the simulation chain. This has the benefit of generating plausible and physically-motivated downstream outcomes that serve as augmentations in a contrastive pre-training, enabling learning of the salient components of physics simulations.
    \item The creation of a large open dataset for common development in the community~\href{https://doi.org/10.5281/zenodo.10633815}{https://doi.org/10.5281/zenodo.10633815}.
    \item A systematic study of the gains of re-simulation-driven contrastive learning against fully-supervised learning strategies. 
\end{itemize}


While the focus of this work will be on jets produced at the LHC, related strategies can be extended to other domains where simulation is present.

\section{Methods}
\label{sec:methods}

\begin{figure*}[t!]
    \centering
    \includegraphics[trim=3.5cm 0cm 10.5cm 0cm, clip, width=0.975\textwidth]{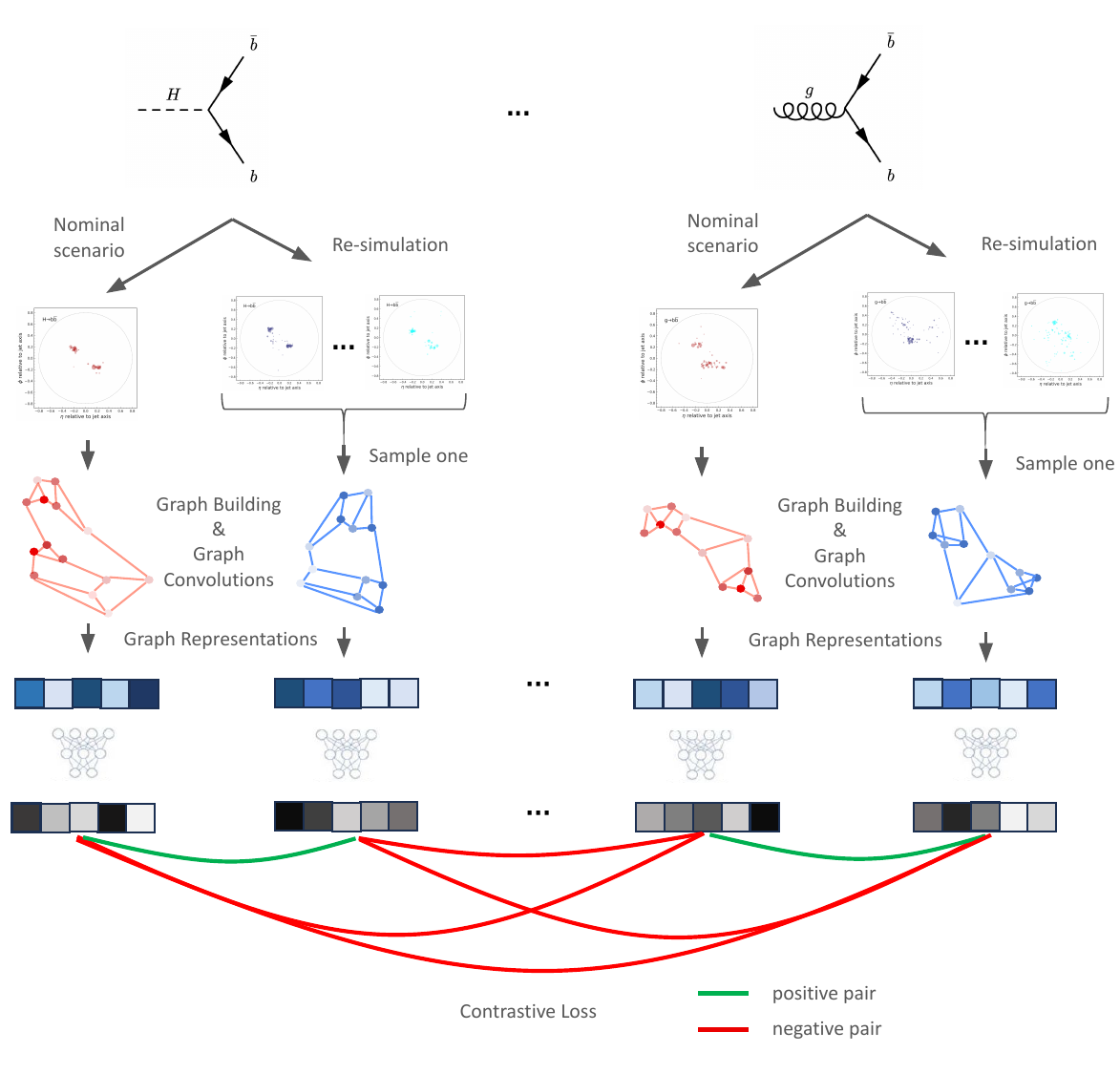}
    \caption{Illustration of the RS3L setup, including downstream re-simulation, sampling, graph computation, and the construction of positive and negative pairs. These are then used in a contrastive loss function aiming to align positive pairs and push negative pairs apart.} 
    \label{fig:everything}
\end{figure*}

Our re-simulation strategy aims to develop a physically motivated pre-training of a model that we dub the RS3L backbone.  This model represents data in a multi-dimensional space rich in information that is common between an object and its augmentations, leading to a localization of the object within the space. Through this strategy, we can further reduce the dimensionality of the overall space, leading to a low-dimensional representation that aims to capture the key properties of the high-level objects.  This strategy is illustrated in~\autoref{fig:everything}, which is described in the following. 

\subsection{The RS3L backbone}

Events are produced by a high-fidelity stochastic simulation of the hard scattering of elementary particles
(in our case, outgoing particles can be Higgs bosons, quarks, or gluons). Different configurations for the parton showering and hadronization are employed during re-simulation, each giving an augmented realization, i.e., a different jet, of the same initial parton. 
In the following, the \textsc{Pythia8}~\cite{Sj_strand_2015} parton shower and hadronization simulation with standardized settings derived from tuning to CMS experimental data~\cite{CMS:2019csb} is used to generate the \emph{nominal} scenario.

Domain completeness can be achieved by considering a set of augmentations which represents the full knowledge embedded in our high-fidelity simulator. Accordingly, we perform augmentations that span an array of plausible experimental outcomes, namely:

\begin{enumerate}
    \item In-domain augmentation by keeping the simulator settings fixed but re-sampling with a different numerical seed
    \item Out-of-domain augmentation by either (i) varying the primary simulator settings within well-motivated bounds, in our case altering the probability for final-state radiation (FSR) branchings in the parton showering step, or (ii) using a different simulator, in our case a different parton shower model (\textsc{Herwig7}~\cite{https://doi.org/10.48550/arxiv.1705.06919}).
\end{enumerate}

The above set captures both 1) implicit uncertainty in the definition of a quark or gluon parton shower, which results in a cascade of particles that can vary in number of particles and distribution of energy across the particles, and 2) the variations within parton shower simulations that are known to cover the observed variations in the data. The augmentations themselves can be interpreted physically as a construction of the parton ``wave function'' before its collapse into a measured state. The resulting space thus aims to ``project away'', i.e., disregard, the imperfect modeling of the parton shower process and the statistical variability in our simulator, including quantum effects and ill-defined hidden parameters.

The RS3L space is trained with 5M events using a 50\%/50\% admixture of QCD jets and Higgs jets. 
Given a nominal jet, we randomly sample one of its augmentations to build a \emph{positive pair}. All other possible jet pairs in a minibatch, which consists of 100 nominal and 100 augmented jets, are considered \emph{negative pairs}.
We then process each jet with the backbone, which we define using graph-based architectures with built-in graph convolutions and message-passing. Graph neural networks (GNNs) have seen large success over other ML algorithms like dense or convolutional neural networks in the analysis of LHC data due to the point-cloud nature of the particles that comprise a jet or an event (for an extensive review, see~\cite{Shlomi_2021}). Finally, each jet is embedded into an $N$-dimensional latent space. We use $N=8$ dimensions for the latent space as a balance between expressiveness and computational complexity, although higher-dimensional spaces are possible\footnote{We observed that networks trained with 128 and 512 output dimensions performed similarly in the downstream tasks without notable improvement over $N=8$.}. The 8D jet representations are then used in the contrastive loss function based on SimCLR. The loss function aims at aligning the 8D vectors of jets building a positive pair, and pushes apart jets from negative pairs. A temperature parameter $\tau$ regulates the relative importance of these two simultaneous training objectives and is set to 0.1. 


\subsection{\label{sec:level2}Data augmentations and training input}

Hard-scattering events for $\text{p}\text{p}\to\text{Z}+\text{jet}$ and $\text{p}\text{p}\to\text{H}\text{Z}$, with $\text{H}\to\text{b}\bar{\text{b}}$ and $\text{Z}\to\nu\bar{\nu}$, from which we sample the jets for training the networks, are generated at a center-of-mass energy of $\sqrt{s}=13\,\text{TeV}$ using \textsc{MadGraph5\_aMC@NLO} v3.4.0~\cite{Alwall_2014} with the NNPDF30\_nlo\_nf\_5\_pdfas parton distribution function~\cite{Ball_2015} at leading-order accuracy in QCD. The j and, respectively, the Higgs boson are generated within a $p_\mathrm{T}$ range of 400-600\,GeV and within $\abs{\eta}<0.1$. The hard-scattering events are then passed through a parton shower simulation and are reconstructed using the \textsc{Delphes} v3.4.3pre1 detector simulation~\cite{delphes} with a CMS detector-like geometry. 
For gluons and the Higgs, the first splitting or, respectively, decay is generated at matrix-element level with a minimum $\Delta R=\sqrt{(\Delta\eta)^2+(\Delta\phi)^2}>0.001$. All further splittings are generated by the subsequent parton shower.

In the nominal configuration, \textsc{Pythia8} v8.244 with the CP5 tune~\cite{CMS:2019csb} is used for parton showering, hadronization, and modelling of the underlying event. No effects from pile-up are simulated.
Four augmentations of each nominal jet are created via re-simulation: (i) the same settings as in the nominal configuration are used, but the partons are passed through \textsc{Pythia8} using a different random initial seed; (ii) the value for the parton shower renormalization scale used in the determination of final-state radiation is multiplied by a factor $\sqrt{2}$; (iii) the same scale is multiplied by a factor $1/\sqrt{2}$;  (iv) \textsc{Herwig7} with the default tune is used for parton showering and hadronization. In all parton shower configurations, initial-state radiation has been disallowed to focus solely radiation effects of the final state.

Jets are finally clustered from the \textsc{Delphes} E-Flow candidates using the anti-$k_t$ algorithm~\cite{Cacciari_2008} with a radius parameter $R=0.8$. The 100 highest-$p_\mathrm{T}$ E-Flow candidates per jet are used for training the networks. For jets to be considered in the training and evaluation, one QCD parton (gluon or quark) or a Higgs boson along with its two decay b quarks have to be matched to the jets, i.e., fulfill $\Delta R<0.8$, with $\Delta\eta$ and $\Delta\phi$ being the difference in pseudorapidity $\eta$ and, respectively, in azimuthal angle $\phi$ between the parton and the jet axis. Jets with a transverse momentum of $p_\mathrm{T}>450$\,GeV, a mass of at least $10\,\mathrm{GeV}$, and $|\eta|<0.1$ are kept. These requirements are also imposed for all augmentations of a given nominal jet. 


\subsection{\label{sec:architecture}Network architecture}

We use a graph neural network that employs a stack of \textsc{DynamicEdgeConv}~\cite{DBLP:journals/corr/abs-1801-07829} network layers for enriching the particle features with information from neighboring particles. The $k$ nearest neighbors are determined dynamically, i.e, in the latent feature space obtained after the initial embedding or, respectively, after each graph convolution. A BERT-like transformer architecture~\cite{DBLP:journals/corr/abs-1810-04805} adapted from~\cite{Maier:2021ymx} has been used to cross-check the results and yields comparable performance. We point out that an optimization of the architecture used for the demonstration of the contrastive approach goes beyond a proof-of-concept and, thus, beyond the scope of this article. The architecture used in the RS3L pre-training is:

\begin{gather}
    h_\mathrm{embed} = \mathrm{MLP}_\mathrm{embed}(X) \nonumber \\ 
    h_\mathrm{DEC1} = \textsc{DynamicEdgeConv}(h_\mathrm{embed}|k=24) \nonumber \\ 
     h_\mathrm{DEC2} = \textsc{DynamicEdgeConv}(h_\mathrm{DEC1}|k=24) \nonumber \\ 
      h_\mathrm{DEC3} = \textsc{DynamicEdgeConv}(h_\mathrm{DEC2}|k=24) \nonumber \\ 
      h_\mathrm{enc} =\mathrm{MLP}_\mathrm{enc}(h_\mathrm{DEC3})
      \nonumber \\ 
    \bm{z} = \textsc{GlobalSumPool}(h_\mathrm{enc}) \ .
\end{gather}

The input feature set $X$ is given by 15 features per particle, listed in Tab.~\ref{tab:inputfeatures}. These input features are not normalized over a minibatch, but their magnitudes are comparable among the features and $\mathcal{O}(1)$, thereby regularizing the network and leading to a similar effect as a pre-transformation that enforces a mean of 0 and standard deviation of 1.

For embedding the 15 particle features into a higher-dimensional latent space, two fully-connected layers with dimensions (128/128) are used in $\textsc{MLP}_\text{embed}$. 
After a series of three convolutions with $k=24$, $\mathrm{MLP}_\mathrm{enc}$ consists of four fully-connected layers with dimensions (64/32/32/8). Outputs of all neurons in both MLPs are activated with an exponential linear unit~\cite{clevert2016accurate}. The final building block, \textsc{GlobalSumPool}, aggregates the eight features of each particle in the jet from the last layer of $\mathrm{MLP}_\mathrm{enc}$ to obtain the final eight features $\bm{z}$ characterizing the jet in the RS3L space.

\begin{table*}
\caption{\label{tab:inputfeatures}%
Description of the particle input features.
}
\begin{center}
\begin{tabular}{r|l}
\textrm{Particle feature}& \textrm{Description}\\
\hline
$\log{p_\text{T}}$ & logarithm of the transverse momentum\\
$\log{(p_\text{T}/p_\text{T}^\text{Jet})}$ & logarithm of $p_\text{T}$ normalized w.r.t. the jet\\
$\log{E}$ & logarithm of the energy\\
$\log{(E/E^\text{Jet})}$ & logarithm of $E$ normalized w.r.t. the jet\\
$\Delta\eta$ & difference in pseudorapidity w.r.t. the jet\\
$\Delta\phi$ & difference in $\phi$ w.r.t. the jet\\
$\Delta R$ & distance in $\eta$-$\phi$ plane from the jet axis\\
$\tanh d_0$ & hyperbolic tangent of the transverse impact parameter\\
$\tanh d_z$ & hyperbolic tangent of the longitudinal impact parameter\\
$q$ & particle charge\\

\texttt{isPhoton} & 1 if particle is a photon, else 0\\
\texttt{isMuon} & 1 if particle is a muon, else 0\\
\texttt{isElectron} & 1 if particle is an electron, else 0\\
\texttt{isCH} & 1 if particle is a charged hadron, else 0\\
\texttt{isNH} & 1 if particle is a neutral hadron, else 0\\
\end{tabular}
\end{center}
\end{table*}

The loss function employed is based on the SimCLR framework introduced in~\cite{DBLP:journals/corr/abs-2002-05709} and has first been used for jet physics in~\cite{10.21468/SciPostPhys.12.6.188}. It is given by

\begin{equation}
    \mathcal{L} = -\log{\frac{e^{s(\bm{z}_i,\bm{z}_i')/\tau}}{\sum_{i\neq j\, \in\, \mathrm{minibatch}}\left[ e^{s(\bm{z}_i,\bm{z}_j)/\tau}+e^{s(\bm{z}_i,\bm{z}_j'
)/\tau}\right]}}
\label{eq:loss}
\end{equation}

with $s(\bm{z}_i,\bm{z}_j)$ being the cosine similarity between negative jet pairs $i$ and $j$ or, in the case of positive pairs, the nominal jet $i$ and the augmented jet $i'$:

\begin{equation}
    s(\bm{z}_i,\bm{z}_j)=\frac{\bm{z}_i\cdot\bm{z}_j}{|\bm{z}_i| |\bm{z}_j|}=\cos\theta_{ij} .
\end{equation}

The network will try to push apart dissimilar jets (negative pairs) on the $N-1$-dimensional hypersphere with unit radius given by the $N$ input features, while trying to align similar jets (positive pairs) and characterize them by similar features. 
The competition between these two effects is regulated by a temperature parameter $\tau$. We studied this parameter based on the performance obtained when using the RS3L space to distinguish Higgs from QCD jets. Specifically, we trained RS3L spaces with $\tau$ ranging from $0.05$ to $0.5$ and fine-tuned these networks on the Higgs vs. QCD classification task with a fixed backbone. A fixed backbone offers the best insight into the performance of the RS3L pre-training step because it projects jets into a one-dimensional classification space using only the information derived during the self-supervised step. We observed that the Higgs vs. QCD binary classification performance was optimal with $\tau = 0.1$. 



\begin{figure}[th]
    \centering
    \includegraphics[width=0.475\textwidth]{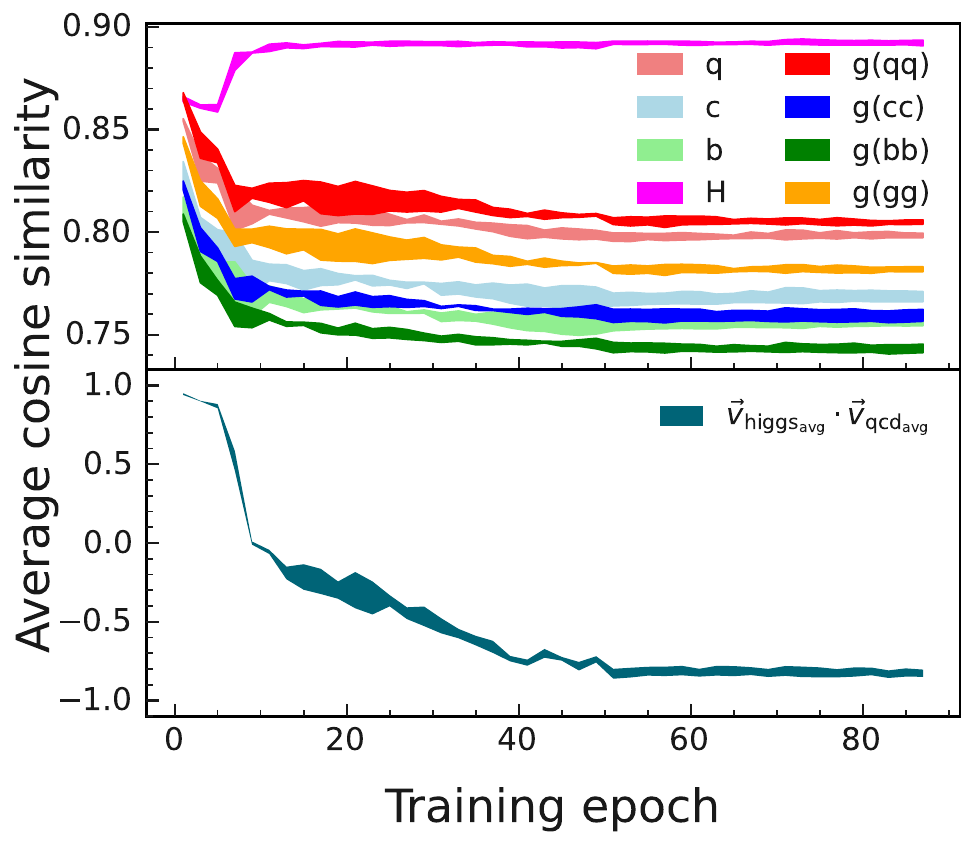} 
    \caption{Metrics pertaining to the convergence of the RS3L training as a function of epoch. Top panel: the average of cosine similarity between the positive pairs (anchor jet and augmented jet). Bottom panel: the cosine similarity between the average Higgs vector and average QCD vector. The variation, indicated by the error bands, is computed over three RS3L trainings.} 
    \label{fig:cosine-similairty}
\end{figure}

\autoref{fig:cosine-similairty} illustrates the convergence of the RS3L space as a function of training time. In the top panel, the cosine similarity is shown for nominal and augmented jets (averaged over each class). This is analogous to the positive pair component of the loss function, the pairs that the network will attempt to make parallel in the space. The network makes augmentations nearly parallel to their augmentations, with average angles between $0.74$ and $0.89$. We observe that the network further aligns Higgs vectors over time, while slightly mis-aligning QCD vectors over time. In the bottom panel, the cosine similarity between the average vectors of the two summary classes (Higgs and QCD) is shown. At the beginning of the training, Higgs and QCD occupy roughly the same region of the space, however as the network converges the populations are pushed to lie on opposing ends of the space, despite not knowing about their different class nature. In all plots, the bands are indicative of averaging over 3 repeated RS3L trainings with random initialization.


\subsection{\label{sec:finetuning}Fine-tuning and fully-supervised trainings}

The MLP placed on top of the RS3L architecture is given by one additional fully-connected layer with 8 input dimensions and a single sigmoid activation function at the output. The backbone network architecture used in the fine-tuning is the exact same as the fully-supervised. The only difference between the approaches is that in the fine-tuning method the base graph trained with RS3L is loaded as a hot-start for the classification and the MLP is added on top of the backbone network. We consider two configurations of fine-tuning the RS3L space: a ``fixed'' graph (where only the MLP parameters are floating) and a ``floating'' graph (where all the MLP and graph parameters are floating). The admixture of QCD/H/W, as well as augmentations composition, are always consistent between the RS3L fine-tuning and full supervision. The RS3L space and classification steps are trained on independent datasets.

In each case, the network is trained repeatedly (at least three times) to gain a measure of whether a result is statistically significant. The numbers in ROCs and tables indicate the averages of these repeated trainings. The difference between subsequent training runs is observed to be quite small.

\section{Results}

\subsection{Understanding the contrastive space}
\label{sec:space}

\begin{figure*}
    \centering
    \includegraphics[width=0.475\textwidth]{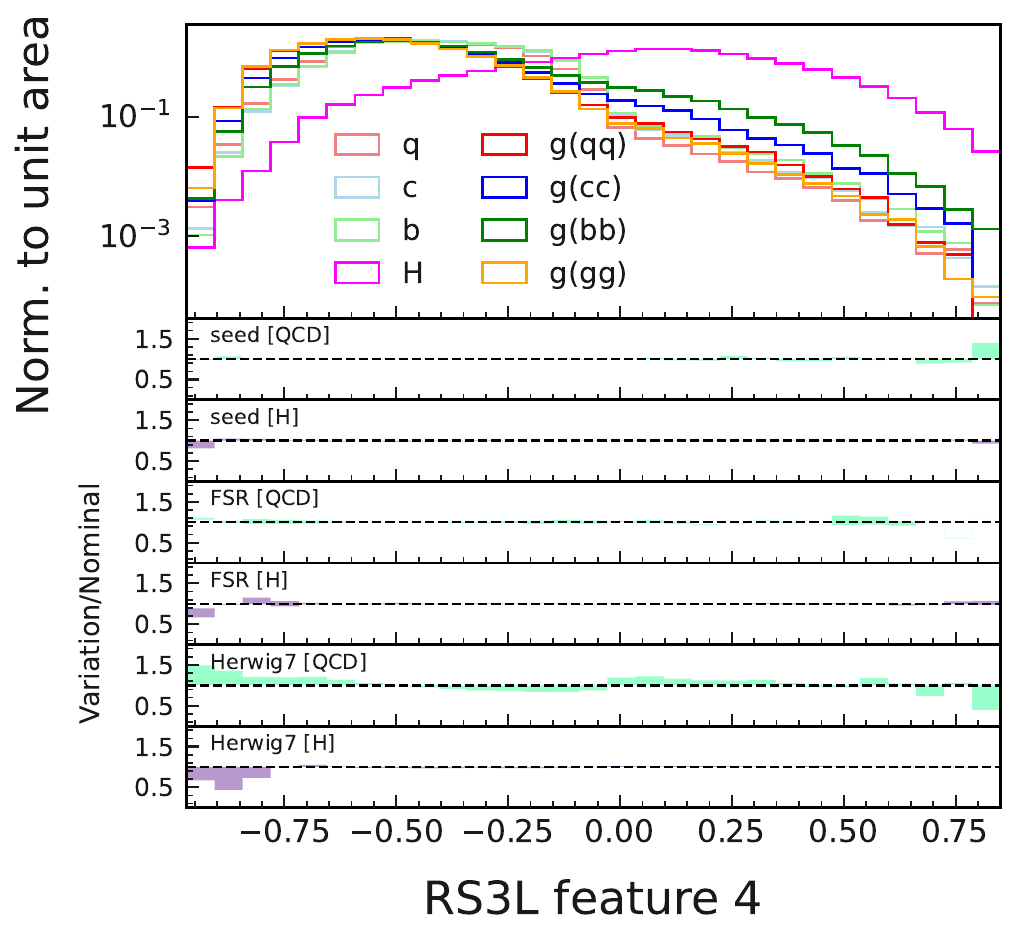}
    \includegraphics[width=0.475\textwidth]{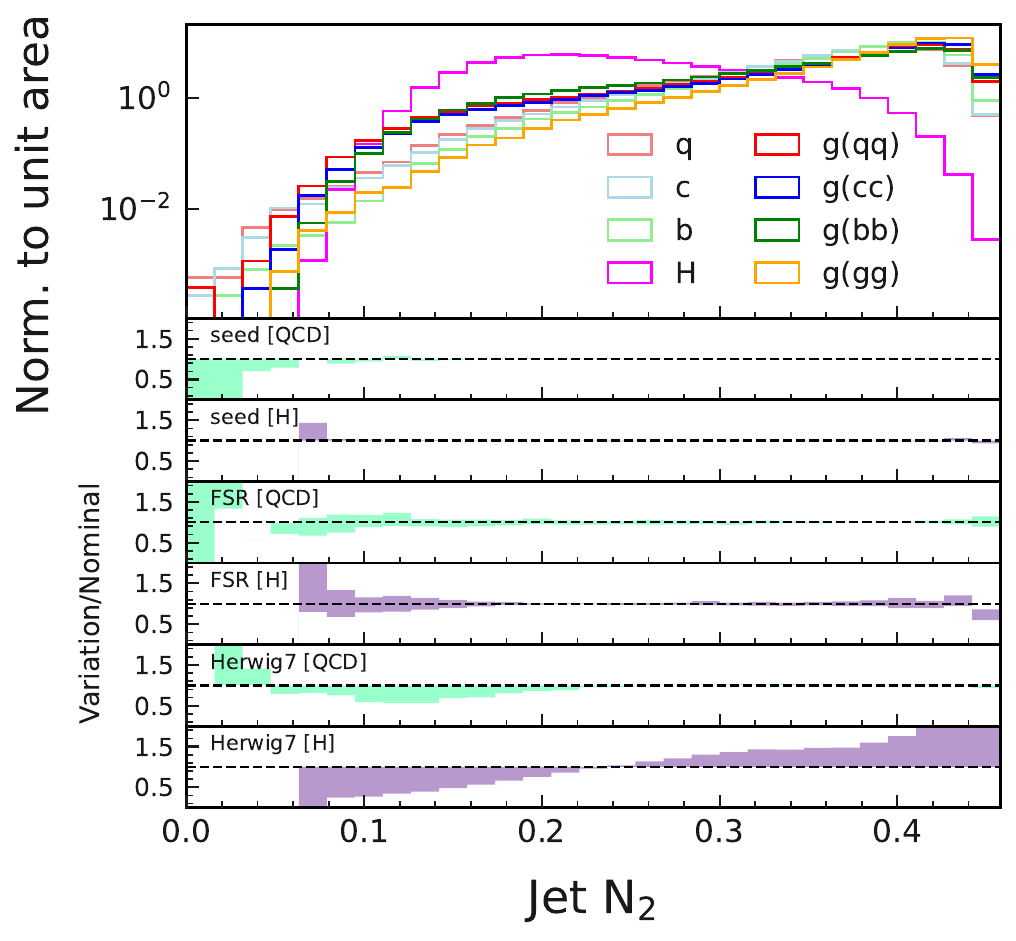}\\
    \caption{(Left) One of eight features derived in the RS3L pre-training. (Right) Jet substructure variable $N_2$. The main (upper) panel shows the distributions for the nominal parton shower scenario. The ratio panels show the difference between the respective varied distributions and the nominal distribution. For FSR, the up and down variations form a band around the nominal distribution.}
    \label{fig:1d-features}
\end{figure*}

Of the eight dimensions, the most discriminating dimension between Higgs bosons jets and quarks and gluons is shown on the left of \autoref{fig:1d-features}. As a qualitative statement on the robustness of the space we derive through RS3L, we also show in the same figure an equivalent plot for the variable $N_2$, which is a theory-motivated function of the energy flow of particles in the jet~\cite{Moult_2016}. The latter represents a ratio between the compatibilities of the jet to have one or, respectively, two prongs, which are regions of collimated radiation. As a powerful jet tagging variable, $N_2$ is commonly used in searches for processes involving hadronically decaying Higgs bosons~\cite{CMS:2018zjv,CMS:2019ykj,CMS:2020zge} or (additional) weak gauge bosons~\cite{CMS:2017dcz,CMS:2019emo} to reduce the overwhelming background where jets arise from quantum chromodynamics (QCD) interactions.
In both plots, the distributions of Higgs jets are shown in magenta and the distributions of QCD jets (jets initiated from quarks and gluons) are shown in other colors (see also \autoref{sec:methods}). In particular, QCD jets are divided by their flavor content and whether the initial parton was a single quark or a gluon. The main (upper) panel shows the RS3L or, respectively, $N_2$ feature for the nominal parton shower scenario, while the bottom panels show the ratios of the varied distributions to the nominal distribution. 
In these plots, ratios closer to 1 indicate smaller differences in the learned representation for jets with different simulator configurations, and therefore smaller systematic uncertainties arising from the parton showering model chosen. Large trends are observed in these ratio when comparing different simulator configurations in the case of $N_2$, while the trends are mitigated in the RS3L space thanks to the employed self-supervision strategy of aligning nominal jets with their augmented versions. 

\begin{figure*}
    \centering
    \includegraphics[width=0.975\textwidth]{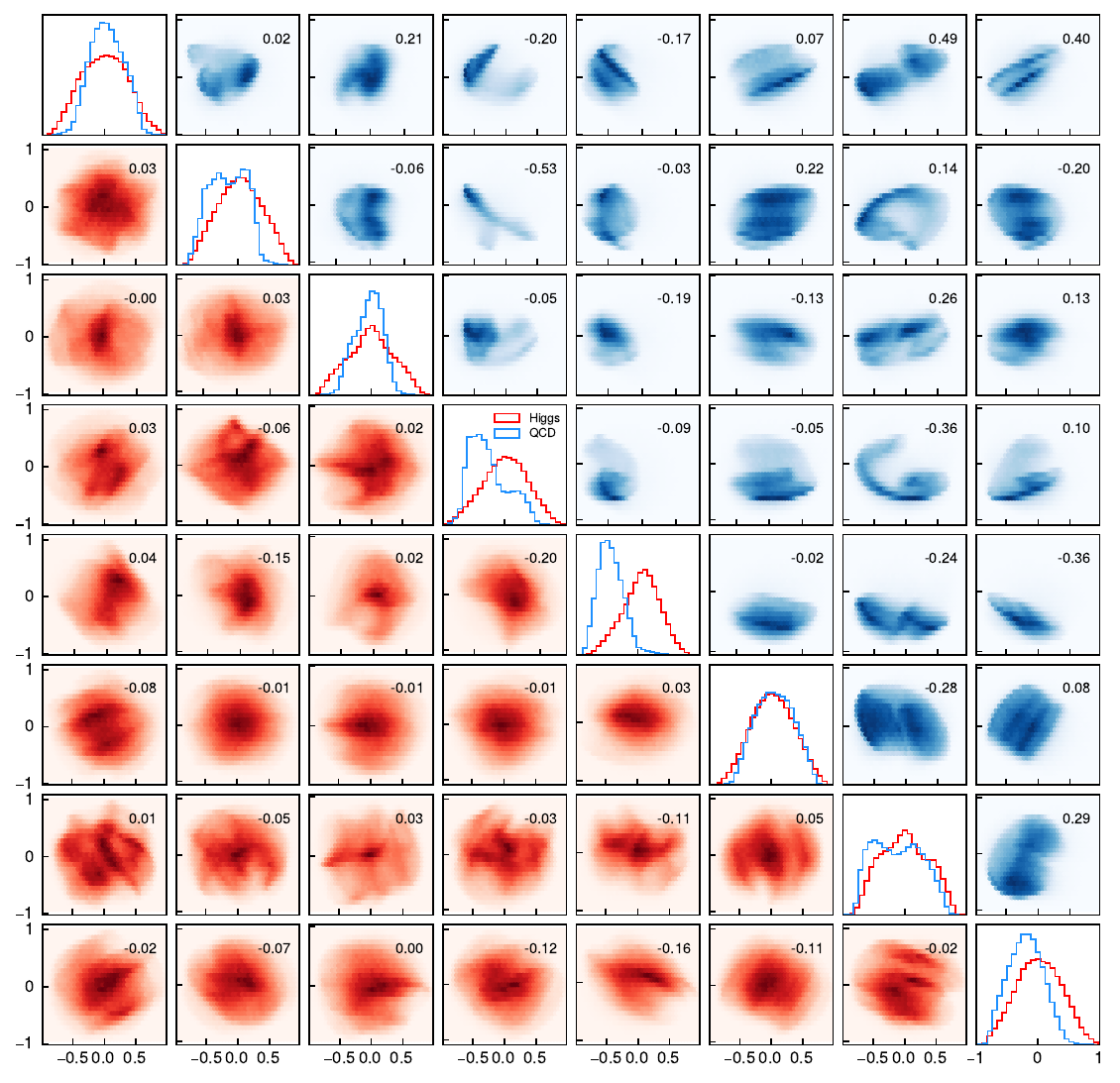}\\
    \caption{Corner plots for the eight outputs of RS3L, split up into Higgs boson (reds) and QCD jets (blues). Only small correlations are observed among feature pairs, as indicated by the Pearson correlation coefficients provided in each subplot.}
    \label{fig:corner}
\end{figure*}

Both the RS3L feature and $N_2$ provide discrimination power between Higgs and QCD jets. This class separation in RS3L comes about despite the network not receiving any information about the true class of the jet. Rather, it emerges dynamically through self-supervision resulting from fulfilling the contrastive training objective.

The corner plots in~\autoref{fig:corner} show the correlation between the eight features derived through RS3L, split up into Higgs and QCD classes. The Pearson correlation coefficient~\cite{pearson} is provided for each feature pair. The largest (anti-)correlation for feature pairs describing Higgs boson jets is $-0.16$, while for QCD jets it is 0.49. Overall, the eight features are largely uncorrelated, indicating a near orthonormal feature set to characterize the jets was found, in particular the ones coming from Higgs bosons.

\begin{figure*}
    \centering
    \includegraphics[trim=0cm 2cm 0cm 0cm, clip, width=0.475\textwidth]{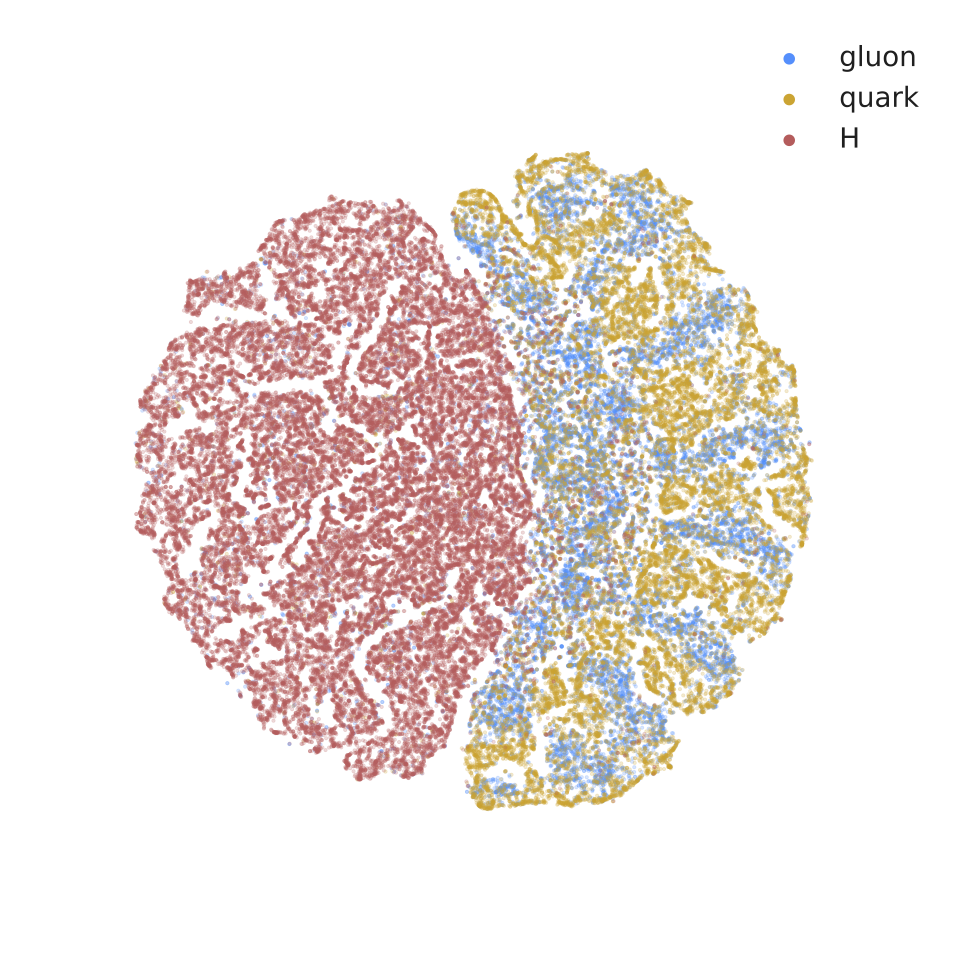}\\
    \includegraphics[trim=0cm 3.4cm 0cm 0cm, clip, width=0.465\textwidth]{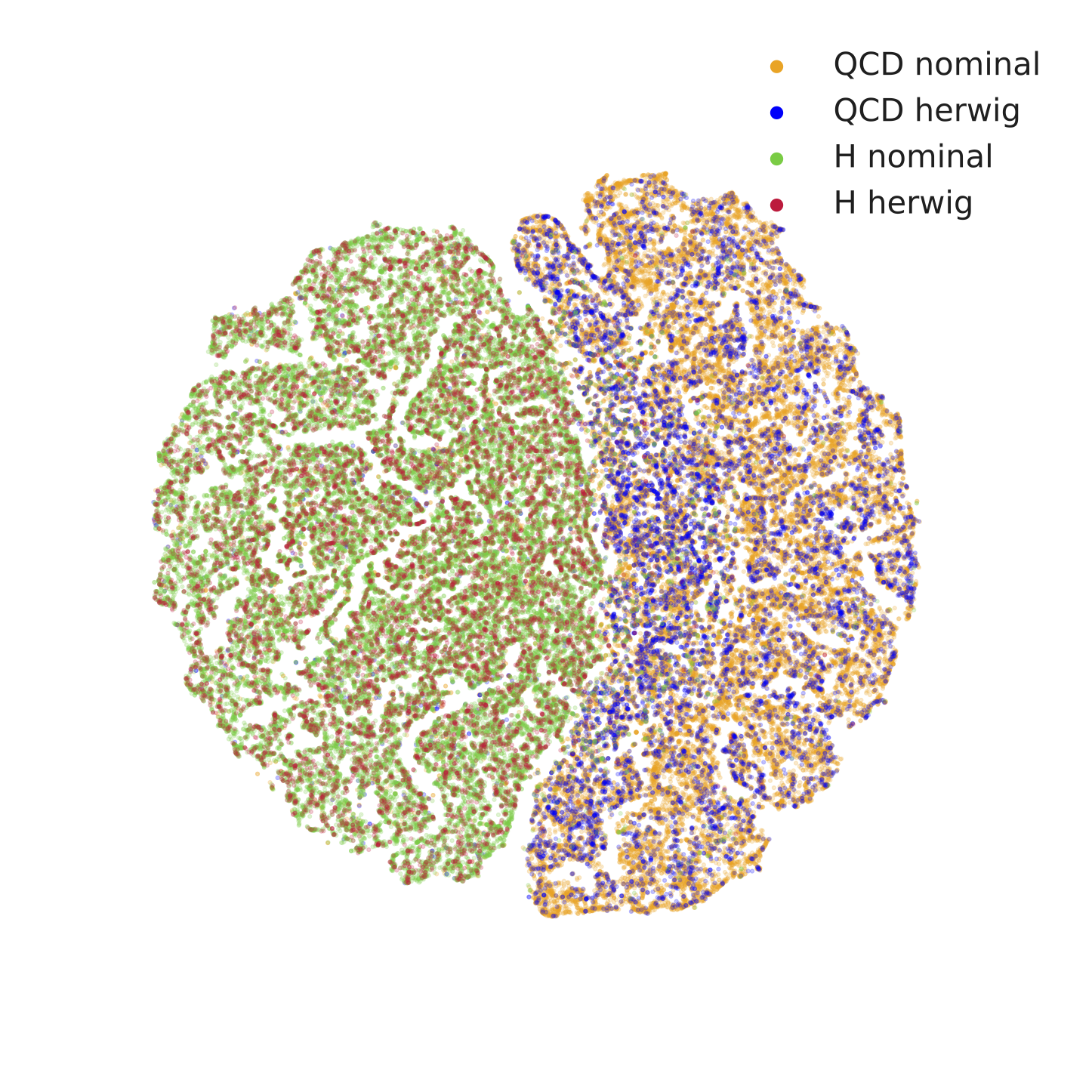}
    \includegraphics[trim=0cm 0cm 0cm 0cm, clip, width=0.435\textwidth]{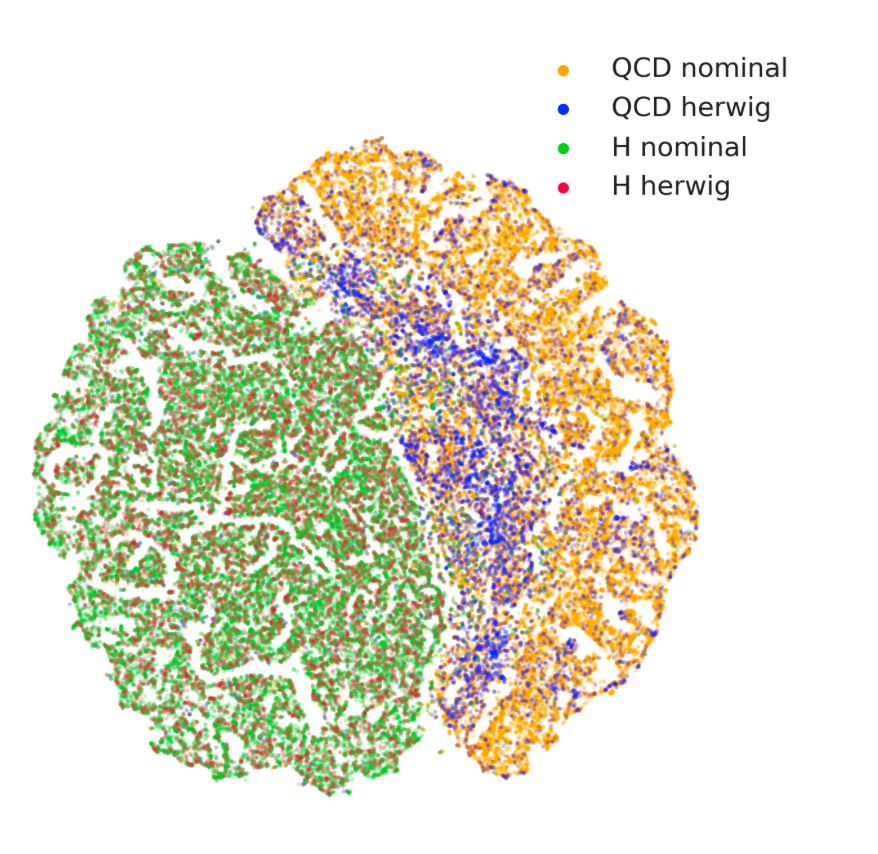}\\
    \caption{2D visualization of the 8D RS3L space, derived via t-SNE dimensionality reduction. Top: A good class separation is seen between Higgs jets and QCD (quark and gluon) jets. Bottom left: Jets shown by parton shower model for \textsc{Pythia8} and \textsc{Herwig7} for a RS3L space trained with \textsc{Herwig7} augmentations. Bottom right: The same for a RS3L space trained without \textsc{Herwig7} augmentations. The congruence of the different parton shower models is visibly worse in the right-hand scenario.}
    \label{fig:tsne}
\end{figure*}

Finally, we visually probe the entire jet representation contained in the 8D RS3L space by performing a reduction to two dimensions using the t-SNE algorithm~\cite{JMLR:v9:vandermaaten08a}. The plot on the top of~\autoref{fig:tsne} shows this 2D space for gluons, quarks, and Higgs jets. Again, strong separation between Higgs and QCD jets as well as strong clustering for a given class is visible. Notably, the model tries to push apart jets from two different Higgs bosons. However, as all Higgs bosons share common characteristics, this objective is hard to meet, leading to the observed clustering. On the bottom left, we show the space divided into jets showered with \textsc{Pythia8} and, respectively, \textsc{Herwig7}. At the population level, it is clear that different showering configurations occupy similar regions in the RS3L space and, hence, our approach provides decent robustness against domain shift. The bottom right of \autoref{fig:tsne} shows the t-SNE 2D space of a RS3L training that did not include \textsc{Herwig7} augmentations. Compared to the RS3L space obtained with all augmentations, this space has visibly smaller overlap (congruence) between \textsc{Pythia8}-showered and \textsc{Herwig7}-showered jets, speaking to larger uncertainties incurred by domain shift.

\subsection{Fine-tuning on top of the RS3L backbone}

We now study the use of RS3L as a backbone for in-distribution and out-of-distribution classification tasks, and we quantify performance and uncertainties. For all fine-tunings, we consider a sample comprising the same types of augmentations used for training the respective RS3L space. We then compare the RS3L pre-training + fine-tuning performance to the performance of a fully-supervised network with the same architecture. For a fair comparison, we also train the fully-supervised network on a sample comprising the same types of augmentations as were used to train the RS3L space. The fully-supervised network, therefore, learns to marginalize over the augmentations, not explicitly exploiting the information that the nominal and augmented jet come from the same initial particle. We point out that the standard approach for fully supervised algorithms at experiments such as ATLAS and CMS is to train on samples comprising only jets from a nominal parton shower configuration.

\subsubsection{In-distribution classification and robustness}

\begin{figure}[t!]
    \centering
        \includegraphics[width=0.475\textwidth]{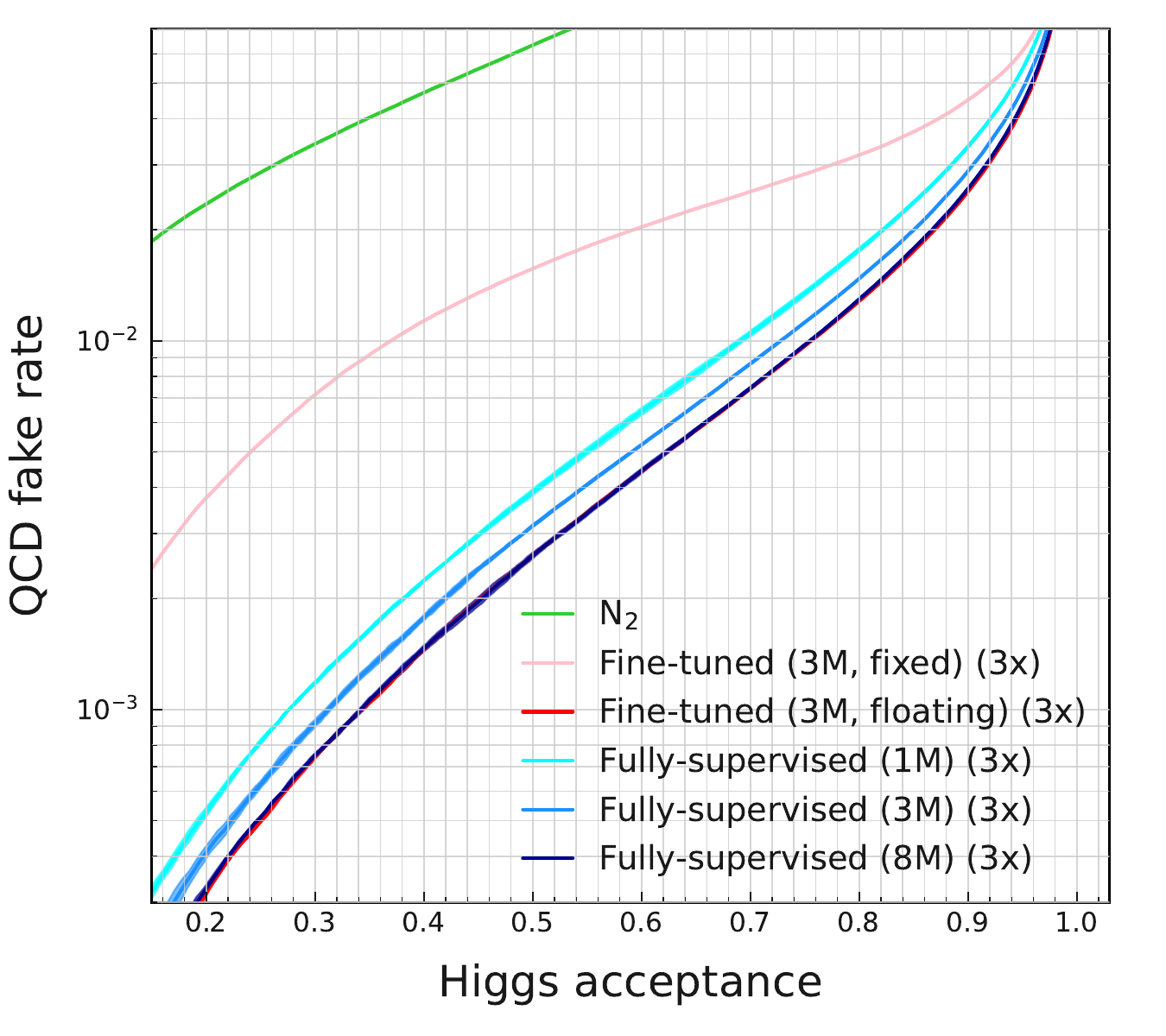} 
    \caption{Tagging performance on the Higgs vs. QCD classification task. All numbers are obtained on a sample comprising only jets from the default parton shower scenario. Fine-tunings on top of a RS3L space that was trained with 5M jets are shown in light and dark red, for graphs that have floating and fixed weights, respectively. The number in parentheses indicates the fine-tuning dataset size. Fully-supervised networks with various training sizes are shown in shades of blue. Three repeated trainings with different random initialization are used to compute an average and standard deviation for each ROC. These are shown as bands. Note that fixed-weight trainings do not have any appreciable uncertainty.}
    \label{fig:roc-HvsQCD}
\end{figure}

\begin{table}[t!]
    \centering
\caption{QCD rejection rates for various training configurations and Higgs efficiencies. The bolded numbers indicate the network with the best performance at a given Higgs efficiency. The uncertainties are calculated using the standard deviation of repeated trainings with different random initializations.}
\vspace{0.2cm}
\begin{tabular}{lrrr}
\toprule
Higgs efficiency          &               0.3 &              0.5 &              0.7 \\
\midrule
$N_2$             &             29 &           16 &            8 \\
Fine-tuned (3M, fixed)    &            $140\pm 1$ &           $64 \pm 0$ &           $39 \pm 0$ \\
Fine-tuned (3M, floating) &  $\mathbf{1325 \pm 31}$ &  $380 \pm 4$ &  $\mathbf{135 \pm 1}$ \\
Fully-supervised (1M)     &   $834 \pm 11$ &  $257 \pm 6$ &   $96 \pm 2$ \\
Fully-supervised (3M)     &  $1070 \pm 29$ &  $317 \pm 3$ &  $115 \pm 1$ \\
Fully-supervised (8M)     &  $1312 \pm 15$ &  $\mathbf{381 \pm 7}$ &  $134 \pm 1$ \\
\bottomrule
\end{tabular}
\label{table:rejection-rates-HvsQCD}
\end{table}

The most straightforward application of RS3L is a binary classification fine-tuning to separate the same jet classes that are used to train the RS3L space. Thus, we first consider the aforementioned canonical classification task of discriminating between jets originating from Higgs bosons and jets originating from QCD. 

Two variants are considered for fine-tuning the RS3L space: (i) The weights of the backbone model are kept fixed, and a task-specific head consisting of a one-layer multilayer perceptron (MLP) is added, whose parameters are trained on the task at hand. (ii) The weights of the backbone model are used as initial parameters but are allowed to be adjusted (``floated'') together with the ones of the task head.

A comparison of RS3L spaces fine-tuned for a Higgs vs. QCD classification task is presented in \autoref{fig:roc-HvsQCD}, which shows the receiver operating characteristic (ROC) curve calculated from \emph{nominal jets only}. The QCD rejection rate (1/ false positive rate) is shown in \autoref{table:rejection-rates-HvsQCD} for various Higgs tagging efficiencies, again computed from only nominal jets. 
Compared to a fully-supervised network trained on 8M jets, a fine-tuning using 3M jets on top of a 5M RS3L space yields a similar performance. If we instead match the yields of the training dataset of the fully-supervised approach and the fine-tuning (i.e., both training on 3M labeled samples), a gain in performance is observed for the RS3L + fine-tuning strategy. This indicates that a common, large pre-training, for which RS3L can be one possible strategy, allows for efficient fine-tuning on comparably much smaller dataset sizes than a fully supervised setup due to the ``hot start''. 

Furthermore, it is important to note that all training configurations that use floating networks are more performant than networks trained with fixed weights on top of the RS3L backbone. This is to be expected because in the fixed weight case the model is severely restricted, as the only trainable parameters belong to the MLP head. Lastly, all networks outperform $N_2$ by a wide margin.

\begin{figure}[th]
    \centering
    \includegraphics[width=0.475\textwidth]{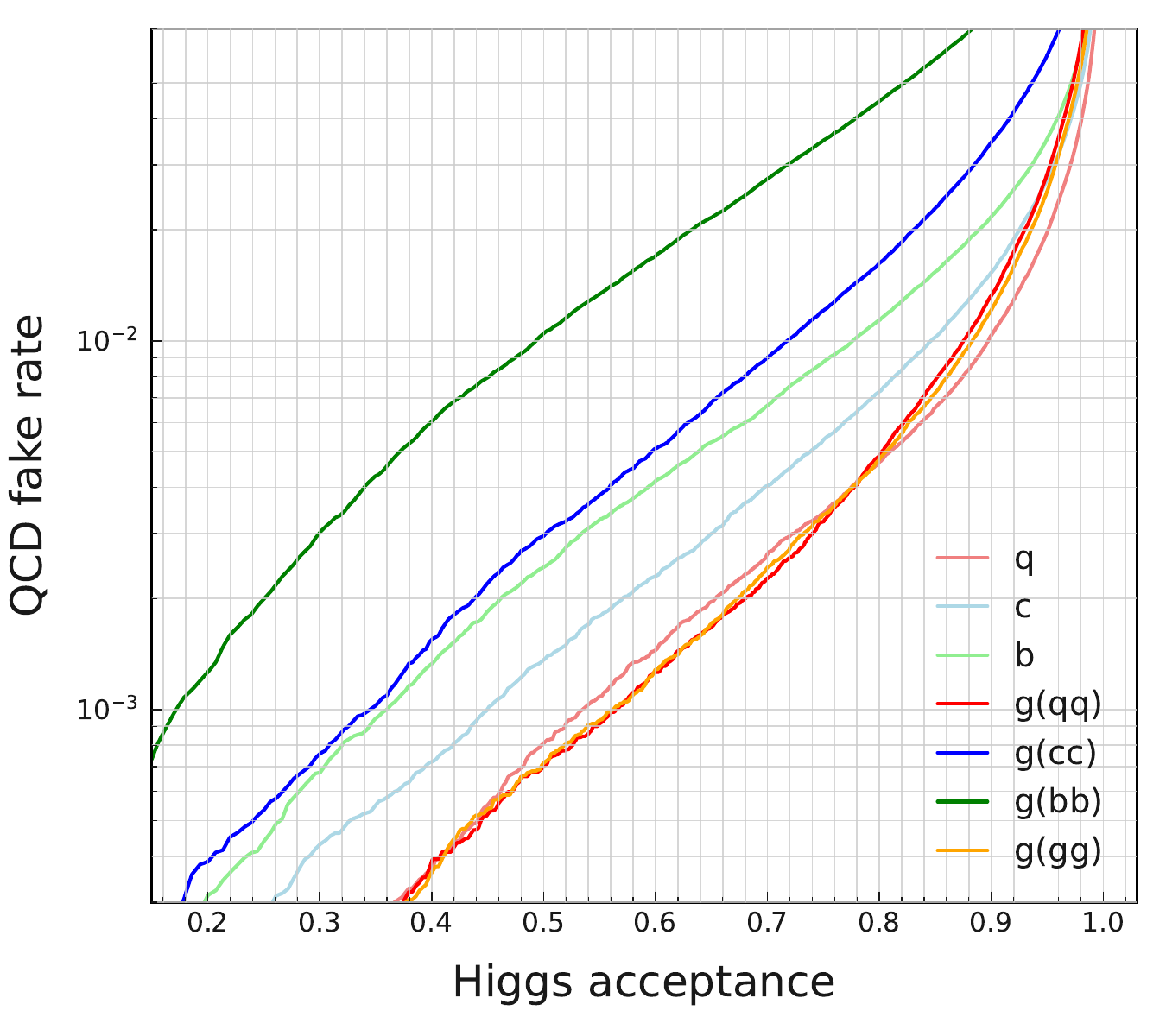} 
    \includegraphics[width=0.475\textwidth]{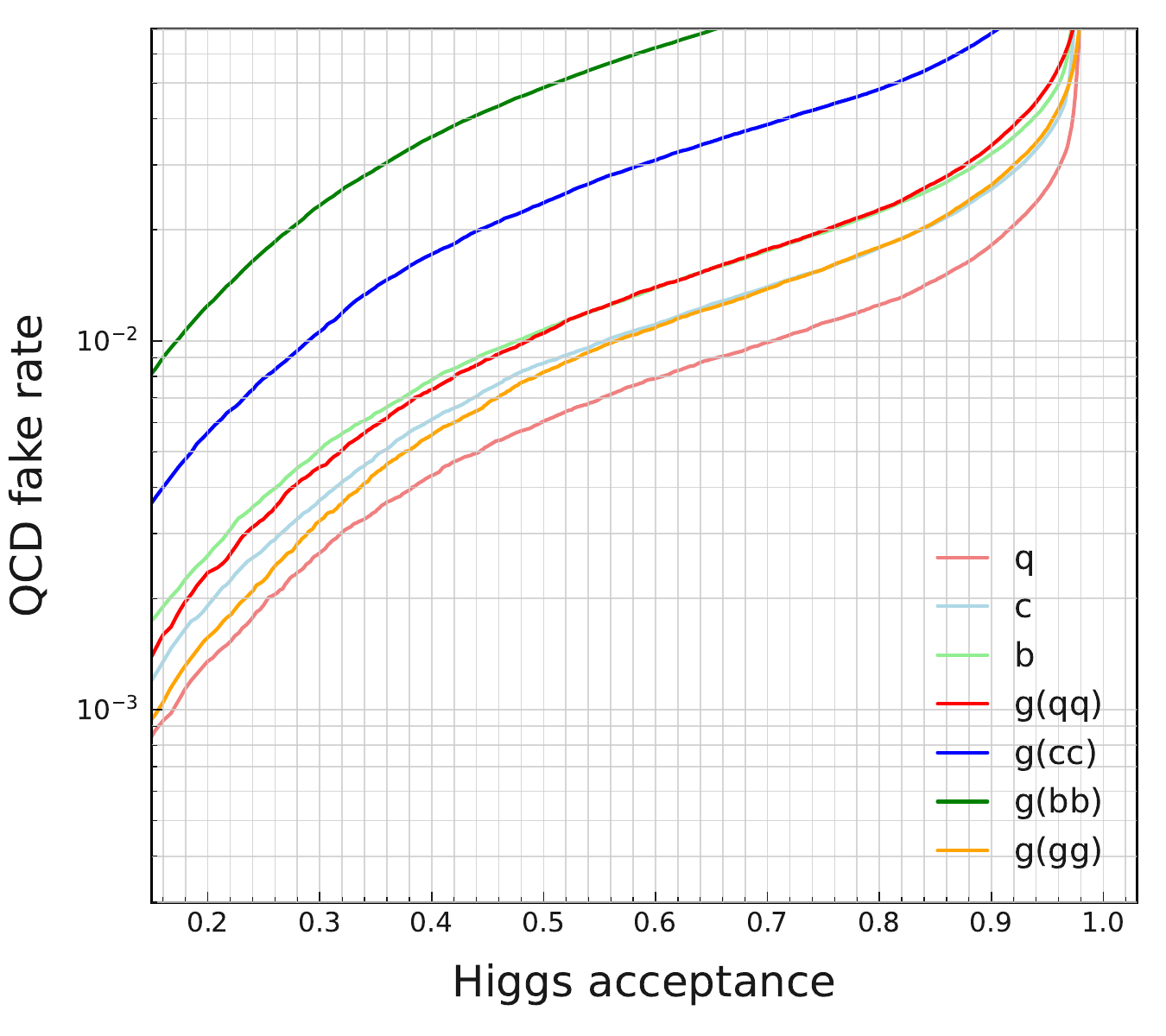} 
    \caption{ROC curves showing the tagging performance of RS3L fine-tunings on the in-domain Higgs vs. QCD classification task. Top: graph fine-tuned with floating weights. Bottom: graph fine-tuned with fixed weights.} 
    \label{fig:ROCs-by-qcdtype}
\end{figure}

The performance of the RS3L space fine-tuned on the in-domain Higgs vs. QCD classification task is shown in ROC curves \autoref{fig:ROCs-by-qcdtype} with separate ROC curves for each individual QCD subprocess. Note that the training sample consists of an admixture with similar weightings of all QCD subprocesses. The top (bottom) shows a graph with floating (fixed) weights. Superior performance is observed for a graph with floating weights in particular for light flavor QCD jets. This is an interesting observation, as it seems that the handles on flavor tagging such as the impact parameters of charged jet constituents get exploited during the fine-tuning step relative to the pre-training stage. For the fine-tuning with floating weights, all light-flavor subprocesses end up with roughly the same ROC curves, while for fine-tuning with fixed weights, the discrimination against single-quark light flavor jets (labelled ``q'') is much better than against light-flavor gluon splittings (labelled ``g(qq)''). In addition, the discrimination power against $\mathrm{g}\to\mathrm{b}\overline{\mathrm{b}}$ to first order does not stem from flavor information, as the Higgs boson also decays into a pair of b quarks. Hence, a direct comparison with the ROC curve for $N_2$ from \autoref{fig:roc-HvsQCD} allows an understanding of the substructure usage, with even fixed-weight RS3L improving over $N_2$ as much as 50\% for a Higgs acceptance of 0.2.

In addition to background rejection, it is necessary to examine the classifier under other metrics. In particular, we consider how the network response varies under different parton shower configurations used in the simulator (which we call the network's ``robustness''). The robustness is quantified by calculating the Wasserstein distance~\cite{ctx54629343510003681,Kantorovich1960MathematicalMO,ramdas2015wasserstein} between the one-dimensional distributions of the classifier output of nominal and, respectively, augmented jets as a measure of the similarity between the responses of the RS3L network to different inputs. The distances are presented in \autoref{table:robustness-combined}.
Each column shows the distance between the distribution for augmented jets (indicated by column title) and nominal jets. A lower Wasserstein distance indicates a network that is less sensitive to the showering configurations. As expected, the distance between nominal jets and jets re-showered with just a different numerical seed is smaller than the distance between nominal jets and any other augmentation (``FSR up'', ``FSR down'', or ``Herwig''). Especially the distance between nominal jets and \textsc{Herwig7}-showered jets is significant. 

Comparing the different training strategies, the most robust networks are trained using the RS3L method. With a fixed-weight approach, the Wasserstein distance is significantly smaller than for all other networks. This is a direct benefit of the RS3L pre-training strategy of minimizing the distance between positive pairs. We note the significant trade-off in terms of worse QCD rejection rates (Table~\autoref{table:rejection-rates-HvsQCD}) for the fixed backbone. 
Including \textsc{Herwig7} augmentations in RS3L or, respectively, full supervision greatly reduces the distances compared to setups without \textsc{Herwig7} augmentations. Interestingly, we observe that already including seed and/or FSR augmentations in RS3L helps fine-tuned networks to be more robust against \textsc{Herwig7}-induced domain shift compared to fully supervised strategies. 
Whether this indicates that a gain in robustness is potentially attainable simply by training the backbone with re-seeded jets merits further investigation that we leave to future studies.


The least robust network is a fully-supervised network trained only on jets with re-seeded configurations (i.e., no systematic variations are seen during training). The deterioration is particularly significant in \textsc{Herwig7}. More robust networks are derived by including all variations in the fine-tuning step. As expected, fully-supervised networks are more robust when the training dataset is increased.


\begin{table*}
\centering
\caption{Wasserstein distances between distributions of tagger output for nominal and augmented jets. Each column indicates the distances between the nominal and the respective augmentation. The number of training events is included for each training setup. Fine-tunings are performed on a 5M RS3L pre-training with either fixed or floating backbone weights. The inclusion of specific augmentations (e.g. ``seed'' or ``FSR'') in the training setup indicates that the pre-training and fine-tuning or supervision is performed on that set of augmentations. Uncertainties are derived from three independent trainings. The bolded numbers indicate the network with the smallest Wasserstein distances.}
\vspace{0.2cm}
\footnotesize
\resizebox{\textwidth}{!}{\begin{tabular}{lllll}
\toprule
             Training setup &                           Seed &                       FSR up &                       FSR down &                         Herwig \\
\midrule
      $N_2$ &            1.51 
           $\times 10^{-4}$ &            2.49 $\times 10^{-3}$ &            2.98 $\times 10^{-3}$ &            1.22 $\times 10^{-2}$ \\
      Fine-tuned (3M, fixed, all) & (8.63 $\pm$ 0.23) $\times 10^{-4}$ & ($\mathbf{4.85 \pm 0.11) \times 10^{-4}}$ & ($\mathbf{1.02 \pm 0.02)\times 10^{-3}}$ &            $\mathbf{(1.08 \pm 0.01)\times 10^{-3}}$ \\
      Fine-tuned (3M, floating, all) & (8.23 $\pm$ 0.24) $\times 10^{-4}$ & (9.09 $\pm$ 0.50) $\times 10^{-4}$ & (1.32 $\pm$ 0.01) $\times 10^{-3}$ & (1.40 $\pm$ 0.04) $\times 10^{-2}$ \\
      Fine-tuned (3M, floating, seed+FSR) & $(\mathbf{7.87 \pm 0.32) \times 10^{-4}}$ & $(9.80 \pm 0.27) \times 10^{-4}$ & $(1.22 \pm 0.02) \times 10^{-3}$ & $(3.58 \pm 0.04) \times 10^{-2}$ \\
      Fine-tuned (3M, floating, seed) & $(8.51 \pm 0.19) \times 10^{-4}$ & $(9.50 \pm 0.18) \times 10^{-4}$ & $(1.36 \pm 0.03) \times 10^{-3}$ & $(3.35 \pm 0.09) \times 10^{-2}$ \\
      Fully-supervised (1M, all) & (9.55 $\pm$ 0.89) $\times 10^{-4}$ & (1.33 $\pm$ 0.04) $\times 10^{-3}$ & (1.45 $\pm$ 0.05) $\times 10^{-3}$ & (2.08 $\pm$ 0.05) $\times 10^{-2}$ \\
      Fully-supervised (3M, all) & (8.87 $\pm$ 0.20) $\times 10^{-4}$ & (1.14 $\pm$ 0.06) $\times 10^{-3}$ & (1.48 $\pm$ 0.09) $\times 10^{-3}$ & (2.18 $\pm$ 0.15) $\times 10^{-2}$ \\
      Fully-supervised (8M, all) & $(8.20 \pm 0.51) \times 10^{-4}$ & $(9.91 \pm 0.33) \times 10^{-4}$ & $(1.29 \pm 0.03) \times 10^{-3}$ & $(1.32 \pm 0.06) \times 10^{-2}$ \\
      Fully-supervised (8M, seed+FSR) & $(8.68 \pm 0.56) \times 10^{-4}$ & $(9.95 \pm 0.06) \times 10^{-4}$ & $(1.18 \pm 0.02) \times 10^{-3}$ & $(3.87 \pm 0.04) \times 10^{-2}$ \\
      Fully-supervised (8M, seed) & $(8.44 \pm 0.15) \times 10^{-4}$ &            $(1.15\pm 0.01) \times 10^{-3}$ & $(1.32 \pm 0.03) \times 10^{-3}$ & $(4.47 \pm 0.32) \times 10^{-2}$ \\

\bottomrule
\end{tabular}}
\label{table:robustness-combined}
\end{table*}

\subsubsection{Out-of-distribution classification task}

We now shift to out-of-distribution tasks to address the degree to which a general representation of showering jets was achieved through the RS3L pre-training. To investigate this, we fine-tune RS3L for a classification task that falls outside of the distributions used for pre-training, namely the task of discriminating between QCD jets and jets from ``hadronic'' W boson decays. Contrary to the Higgs boson decays, jets from W boson decays do not feature $B$ hadrons\footnote{The authors note that the absolute tagging performance on the W vs. QCD and Higgs vs. QCD datasets cannot be compared directly because of the composition of the sample, see \autoref{sec:methods}.}. The experimental signature therefore falls out of the distribution of the Higgs jets used to train the RS3L backbone. 

We follow the same procedure as for the in-distribution classification studies by considering the same pre-trained RS3L backbone but now fine-tuned for W vs. QCD classification. We compare with networks fully-supervised from scratch for W vs. QCD classification. The results of this study are shown in \autoref{table:rejection-rates-WvsQCD}. Networks fine-tuned show significant improvement (as much as 10\%) in background rejection over fully-supervised networks when trained on the same number of examples. 

\begin{figure}[ht]
    \centering
    \includegraphics[width=0.475\textwidth]{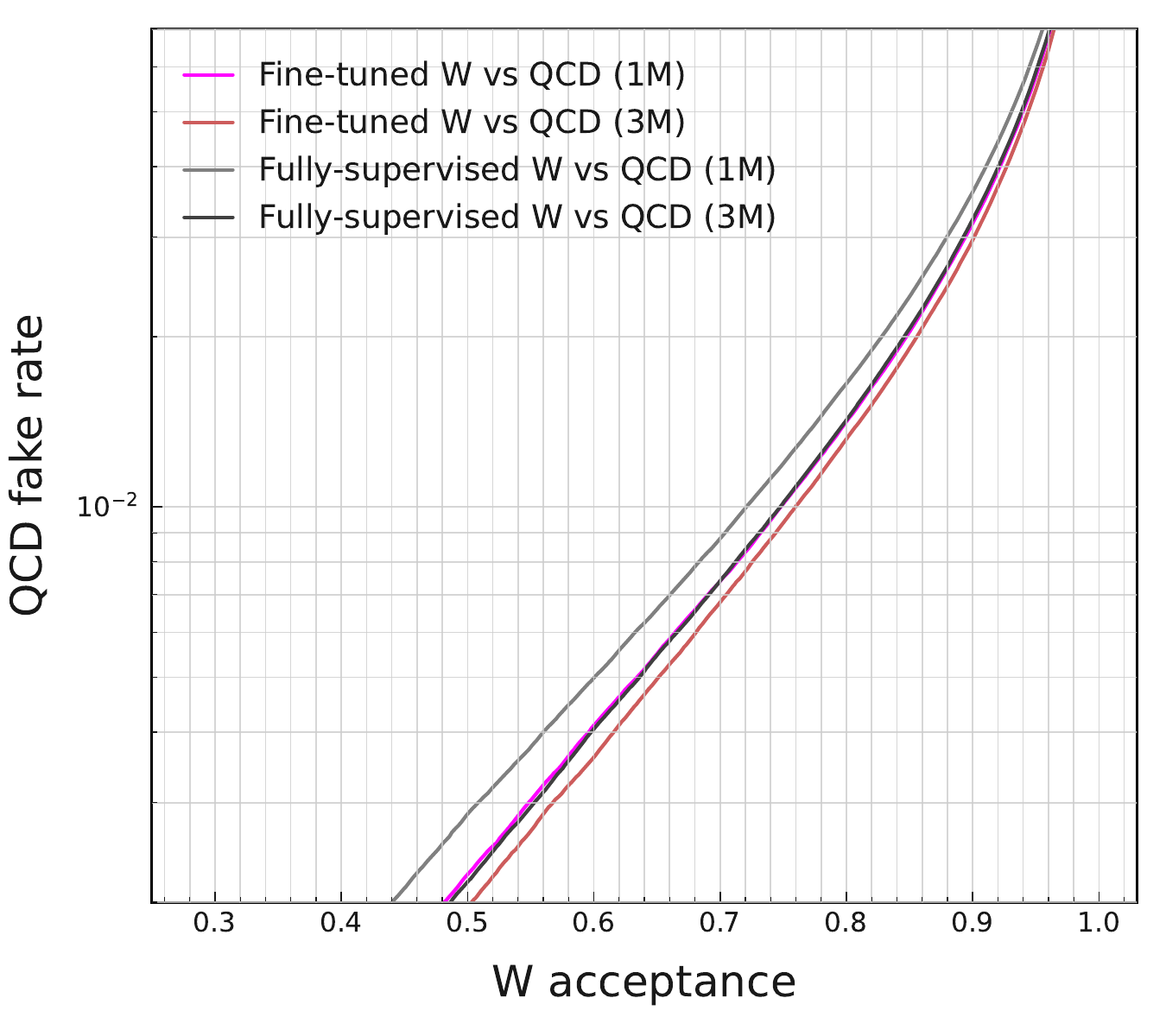} 
    \caption{Tagging performance on the W vs. QCD classification task. RS3L spaces that are fine-tuned for the task are shown in red. Fully-supervised networks with various dataset training sizes are shown in shades of grey. }
    \label{fig:roc-WvsQCD}
\end{figure}

In terms of robustness, the Wasserstein distances between nominal and augmented jets are shown in terms of QCD background rejection in \autoref{table:robustness-combined-WvsQCD} and as a ROC in \autoref{fig:roc-WvsQCD}. The training setups with RS3L are found to be significantly more robust.

\begin{table}[t!]
\centering
\caption{QCD rejection rates for various training configurations and W efficiencies. The bolded numbers indicate the network with the best performance at a given W efficiency.}
\vspace{0.2cm}
\begin{tabular}{lrrr}
\toprule
W efficiency          &               0.3 &              0.5 &              0.7 \\
\midrule
Training setup            & \multicolumn{3}{c}{1/(QCD efficiency)}                    \\ 
\midrule
Fine-tuned (1M, floating, all)       &   1589 $\pm$ 88 &   438 $\pm$ 9 &  134 $\pm$ 2 \\
Fine-tuned (3M, floating, all)       &   $\mathbf{1928 \pm 31}$ &   $\mathbf{504 \pm 3}$ &  $\mathbf{147 \pm 1}$ \\
Fully-supervised (1M, all) &  1288 $\pm$ 102 &  357 $\pm$ 16 &  114 $\pm$ 2 \\
Fully-supervised (3M, all) &   1763 $\pm$ 18 &   459 $\pm$ 1 &  137 $\pm$ 2 \\
\bottomrule
\end{tabular}
\label{table:rejection-rates-WvsQCD}
\end{table}

\begin{table*}
\centering
\caption{Wasserstein distances between distributions of tagger output for nominal and augmented jets in the out-of-domain case. Each column indicates the distances between the nominal and the respective augmentation. The distance is calculated over a sample comprising equal amounts of W and QCD jets. The number of training events is included for each training setup. Finally, uncertainties are derived through measuring the Wasserstein distance for repeated trainings using the same configuration with random initialization.}
\vspace{0.2cm}
\footnotesize
\resizebox{\textwidth}{!}{\begin{tabular}{lllll}
\toprule
             Training setup &                           Seed &                       FSR up &                       FSR down &                         Herwig \\
             \midrule
            Fine-tuned W vs QCD (1M, floating, all) & $\mathbf{(2.48 \pm 0.95) \times 10^{-4}}$ & $(5.18 \pm 0.03) \times 10^{-3}$ &            $(6.53 \pm 0.01) \times 10^{-3}$ & $(4.07 \pm 0.05) \times 10^{-2}$ \\
            Fine-tuned W vs QCD (3M, floating, all) & $(3.26 \pm 0.24) \times 10^{-4}$ & $\mathbf{(4.79 \pm 0.09) \times 10^{-3}}$ &            $\mathbf{(6.08 \pm 0.01) \times 10^{-3}}$ & $\mathbf{(3.58 \pm 0.06) \times 10^{-2}}$ \\
            Fully-supervised W vs QCD (1M, all) & $(3.09 \pm 0.39) \times 10^{-4}$ & $(5.02 \pm 0.19) \times 10^{-3}$ & $(6.28 \pm 0.32) \times 10^{-3}$ & $(4.51 \pm 0.39) \times 10^{-2}$ \\
            Fully-supervised W vs QCD (3M, all) & $(4.08 \pm 0.24) \times 10^{-4}$ &            $(5.22 \pm 0.01) \times 10^{-3}$ & $(6.70 \pm 0.13) \times 10^{-3}$ & $(4.33 \pm 0.19) \times 10^{-2}$ \\
\bottomrule
\end{tabular}}
\label{table:robustness-combined-WvsQCD}
\end{table*}


\section{Outlook}
\label{sec:concl}

RS3L is a novel strategy that combines the concept of re-simulation with a contrastive loss function to drive self-supervised representation learning. The framework provides a natural way for creating a powerful foundation model which can be used for various downstream tasks such as classification and uncertainty mitigation. By encapsulating systematic uncertainties as well as the stochastic variability of the simulation into data augmentations, a latent space with improved robustness compared to other learning strategies can be obtained. 

RS3L has been successfully applied to the canonical task of jet tagging, which is the ability to identify the type of elementary particle at the origin of the evolution of a particle shower in detectors at the Large Hadron Collider. Here, the algorithm revealed identical performance in the limit of large-enough training statistics to state-of-the-art deep learning-based jet taggers that separate jets from Higgs bosons and from QCD partons, improved performance at smaller dataset sizes, and improved robustness against detrimental effects arising from systematic uncertainties. Additionally, it shows excellent transferrability to out-of-distribution tasks, thus having large potential to increase the efficiency of deep learning trainings in high energy physics when employed as a common pre-training.


Enforcing domain completeness through a strategy to embed the known modeling uncertainties within the self-supervised space directly maps this work to the quality of the simulation used at training. Improved simulators can lead to improved embeddings, which, in turn, lead to a better understanding of the underlying physics. As a result, RS3L can adapt to the next generation of simulation modeling that stands to be substantially better than the current simulation. Through self-supervised approaches, we can continually improve the quality of the physics that is extracted, provided we continue to improve the physics model in the samples used for the self-supervision. 

In future work, alternative self-supervised learning approaches to the employed SimCLR strategy, such as VICReg~\cite{DBLP:journals/corr/abs-2105-04906}, SimSiam~\cite{DBLP:journals/corr/abs-2011-10566}, and BarlowTwins~\cite{DBLP:journals/corr/abs-2103-03230} should be explored to study the practical impact of precise components of their loss functions. Additionally, the dataset size for the pre-training should be varied to systematically study the regimes in which the performances of SSL, fine-tuning, and fully-supervised strategies may saturate. Finally, detailed comparisons should be made with other pre-training strategies proposed for HEP, such as masked particle modelling.

\section*{Data Availability}

The RS3L dataset is published at~\href{https://doi.org/10.5281/zenodo.10633815}{https://doi.org/10.5281/zenodo.10633815}. 

\section{Acknowledgments}
The authors would like to thank M. Pierini and S. Mishra-Sharma for fruitful discussions. The authors thank the CERN storage team for the creation and maintenance of a dedicated storage space for this project. The trainings for this study have been performed on the MIT Satori and subMIT clusters. 

BM acknowledges the support of the Alexander von Humboldt foundation and of Schmidt Sciences. MK and JK are supported by the US Department of Energy (DOE) under grant DE-AC02-76SF00515. PH and JK are supported by the Institute for Artificial Intelligence and Fundamental Interactions (IAIFI) under the NSF grant \#PHY-2019786 and the Accelerated AI Algorithms for Data Driven Discovery Grant (A3D3) under NSF grant \#PHY-2117997. 


\bibliography{references}

\end{document}